\documentclass{article}
\usepackage{graphicx}
\usepackage{amsmath}
\usepackage{amsfonts}
\usepackage{amssymb}
\begin{document}

\begin{center}
A study on the\ coexistence of BEC and BCS states

C. D. Hu

Department of Physics National Taiwan University, Taipei, Taiwan R. O. C.

\bigskip

Abstract

\medskip
\end{center}

We pointed out in this work that in dealing with the BEC-BCS crossover
problem, the dynamic effect of
%TCIMACRO{\TEXTsymbol{<}}%
%BeginExpansion
$<$%
%EndExpansion
b(t)%
%TCIMACRO{\TEXTsymbol{>}}%
%BeginExpansion
$>$%
%EndExpansion
is not negligible. Accordingly, an equation of motion approach was devised to
calculate the Green$^{\prime}$s functions. Based on our result, we concluded
that instead of crossover, BCS states and Bose-Einstein condensation always coexist.


\pagebreak 

\noindent

The physics of BCS states in a dilute system was first studied by Eagles[1]
who proposed the possibility of forming Cooper pairs before the onset of
superconductivity. The BEC-BCS crossover problem was discussed by Leggett[2]
in 1980. He pointed out that there is no distinct boundary between Cooper
pairs and Bose-Einstein condensation. As elaborated later by Nozi\'{e}res and
Schmitt-Rink[3],\ in the strong coupling or low density limit, fermions will
combine to form bosonic molecules and condense at sufficiently low
temperature. As the coupling is weakened or density increases, the molecular
wave functions overlap. With sufficient overlapping, the molecular BEC will be
transformed into BCS states. The authors [2,3] saw no phase transition. In
their opinion, the amplitudes of the states of phase coherence (BEC and BCS)
undergo smooth variation.

On the other hand, there has been marvellous experimental progress[4-8]
recently. Fermionic atoms in a trap can form bosonic molecules via Feshbach
resonance[9]. The coupling strength can be adjusted by changing the applied
magnetic field. The energy gap was found on both sides of resonance and its
spectrum is compatible with theoretical calculation[10]. Hence, the scenario
envisaged by Leggett[2] and Nozi\'{e}res and Schmitt-Rink[3] seems to be
realized, with one complication: the fermions and bosons have internal degrees
of freedom. They are the spins of electrons and nuclei. The above mentioned
Feshbach resonance is fulfilled with the aid of hyperfine interaction. As a
result, the state of electron spins of the atoms in the open channel are
different, in general, from those in the closed channel.

Much theoretical work has been devoted to this subject. Kokkelmans et. al.[11]
made a thorough analysis of Feshbach resonance of the cold atom system. Their
deduction greatly simplified further computation. Ohashi and Griffin[12]
adopted the orthodox Feynman diagram approach. Stajic et' al.[13] applied the
concept of pseudogap originated from high $T_{C}$ superconductivity to explain
the experimental results. In this paper we would like to raise a neglected
point which has profound effect on one's understanding of the BEC-BCS system.

We started with the of fermion-boson Hamiltonian[14]:
\begin{align}
H  &  =\sum_{\mathbf{p,}\sigma}\varepsilon_{\mathbf{p}}c_{\mathbf{p,}\sigma
}^{\dagger}c_{\mathbf{p,}\sigma}+\sum_{\mathbf{q}}(\omega_{\mathbf{q}}%
+\nu)b_{\mathbf{q}}^{\dagger}b_{\mathbf{q}}-U\sum_{\mathbf{k,p,q}%
}c_{\mathbf{p+q,\uparrow}}^{\dagger}c_{\mathbf{k-q,\downarrow}}^{\dagger
}c_{\mathbf{k,\downarrow}}c_{\mathbf{p,\uparrow}}\nonumber\\
&  +g\sum_{\mathbf{p,q}}b_{\mathbf{q}}^{\dagger}c_{-\mathbf{p+q/2,\downarrow}%
}c_{\mathbf{p+q}/2,\uparrow}+H.c. \tag{1}%
\end{align}
where $c_{\mathbf{p,}\sigma}$ and $b_{\mathbf{q}}$ are the fermion and boson
operators, $\varepsilon_{\mathbf{p}}$\ and $\omega_{\mathbf{q}}$ are their
energies, $\nu$ is the detuned energy of bosons, and $\sigma$ is the spin
index. The terms in the second line denote the combination of atoms and
dissociation of molecules. They are necessary because, as we stated before,
the states of spins had been changed. Cooper pairs have different spin states
from those in BEC. So they should be treated as different species.

Now we calculate the Green's functions $G_{\sigma}(\tau,\tau^{\prime
};\mathbf{k})=-\langle\widehat{\mathbf{T}}c_{\mathbf{k,\sigma}}(\tau
)c_{\mathbf{k},\sigma}^{\dagger}(\tau^{\prime})\rangle$ and $F_{\alpha\beta
}^{\ast}(\tau,\tau^{\prime};\mathbf{k)=}\langle\widehat{\mathbf{T}%
}c_{\mathbf{k,\alpha}}^{\dagger}(\tau)c_{\mathbf{k},\beta}^{\dagger}%
(\tau^{\prime})\rangle$ where $\widehat{\mathbf{T}}$ is the time-order
operator. The approach of equations of motion was applied.
\begin{align}
\frac{d}{d\tau}G_{\uparrow}(\tau,\tau^{\prime};\mathbf{k})  &  =-\delta
(\tau-\tau^{\prime})-\varepsilon_{\mathbf{p}}G_{\uparrow}(\tau,\tau^{\prime
};\mathbf{k})-[UF_{\downarrow\uparrow}(\tau,\tau)-g\langle b_{0}%
\rangle]F_{\downarrow\uparrow}^{\ast}(\tau,\tau^{\prime};\mathbf{k}%
)\nonumber\\
&  +g\sum_{\mathbf{q}}\langle\mathbf{T}[b_{\mathbf{q}}(\tau)-\langle
b_{0}\rangle]c_{-\mathbf{k+q,\downarrow}}^{\dagger}(\tau)c_{\mathbf{k}%
,\uparrow}^{\dagger}(\tau^{\prime})\rangle\tag{2}%
\end{align}
and
\begin{align}
\frac{d}{d\tau}F_{\uparrow\downarrow}^{\ast}(\tau,\tau^{\prime};\mathbf{k})
&  =\varepsilon_{\mathbf{p}}F_{\uparrow\downarrow}^{\ast}(\tau,\tau^{\prime
};\mathbf{k})+[UF_{\uparrow\downarrow}^{\ast}(\tau,\tau)-g\langle b_{0}%
\rangle]G_{\downarrow}(\tau,\tau^{\prime};\mathbf{k})\nonumber\\
&  +g\sum_{\mathbf{q,}\sigma}\langle\widehat{\mathbf{T}}[b_{\mathbf{q}%
}^{\dagger}(\tau)-\langle b_{0}^{\dagger}\rangle]c_{\mathbf{k+q,\downarrow}%
}(\tau)c_{\mathbf{k},\downarrow}^{\dagger}(\tau^{\prime})\rangle\tag{3}%
\end{align}
where $F_{\alpha\beta}(\tau,\tau)=\sum_{\mathbf{p}}\langle
c_{-\mathbf{p,\alpha}}(\tau)c_{\mathbf{p},\beta}(\tau)\rangle$. Here we
assumed the Hartree-Fock terms can be absorbed into the single-particle
energy. A term $\langle b_{\mathbf{q=}0}\rangle$ had been intentionally added
to and substracted from the right hand side to facilitate later decoupling.

While developing superconductivity theory, $F_{\alpha\beta}(\tau,\tau)$ is
usually set to be a real constant independent of $\tau$. The prescription is
to subtract $N\mu$ from $H$[15]. Similarly, in calculation involving BEC,
$\langle b_{0}(\tau)\rangle$ is also treated as a constant. In the case when
both $F_{\alpha\beta}(\tau,\tau)$ and $\langle b_{0}(\tau)\rangle$ are
present, this method in general, can only make either one of them a constant
but not both. This point was ignored by the preceding works on the BEC-BCS
crossover problem. In this paper, we chose to let $F_{\alpha\beta}(\tau
,\tau)=F_{\alpha\beta}$ be a constant and $\langle b_{0}(\tau)\rangle$ a
function of $\tau$. This is the gist of this work.

With $H$ replaced by $H-N\mu$, we have to substitute $\xi_{\mathbf{k}}%
\equiv\varepsilon_{\mathbf{k}}-\mu$ for $\varepsilon_{\mathbf{k}}$ and
$\nu^{\prime}=\nu-2\mu$ for $\nu$ in eqs. (1-3). If $\langle b_{0}%
(\tau)\rangle$ was treated dynamically, we arrived at the next set of
equations of motion:
\begin{align}
&  \quad\frac{d}{d\tau}\langle\widehat{\mathbf{T}}[b_{\mathbf{q}}%
(\tau)-\langle b_{0}\rangle]c_{-\mathbf{k+q,\downarrow}}^{\dagger}%
(\tau)c_{\mathbf{k},\uparrow}^{\dagger}(\tau^{\prime})\rangle\nonumber\\
&  =[-(\omega_{\mathbf{q}}+\nu^{\prime})+\xi_{\mathbf{k-q}}]\langle
\widehat{\mathbf{T}}[b_{\mathbf{q}}(\tau)-\langle b_{0}\rangle
]c_{-\mathbf{k+q,\downarrow}}^{\dagger}(\tau)c_{\mathbf{k},\uparrow}^{\dagger
}(\tau^{\prime})\rangle-(\omega_{\mathbf{q}}+\nu^{\prime})\langle b_{0}%
\rangle\langle\widehat{\mathbf{T}}c_{-\mathbf{k+q,\downarrow}}^{\dagger}%
(\tau)c_{\mathbf{k},\uparrow}^{\dagger}(\tau^{\prime})\rangle\nonumber\\
&  +U\sum_{\mathbf{p,q}^{\prime}}\langle\widehat{\mathbf{T}}[b_{\mathbf{q}%
}(\tau)-\langle b_{0}\rangle]c_{\mathbf{p+q}^{\prime},\uparrow}^{\dagger}%
(\tau)c_{-\mathbf{k+q-q}^{\prime}\mathbf{,}\downarrow}^{\dagger}%
(\tau)c_{\mathbf{p,\uparrow}}(\tau)c_{\mathbf{k},\uparrow}^{\dagger}%
(\tau^{\prime})\rangle\nonumber\\
&  -g\sum_{\mathbf{p}}\{\langle\widehat{\mathbf{T}}[b_{\mathbf{q}}%
(\tau)-\langle b_{0}\rangle]b_{\mathbf{p}}^{\dagger}(\tau)c_{\mathbf{k-q+p}%
,\uparrow}(\tau)c_{\mathbf{k},\uparrow}^{\dagger}(\tau^{\prime})\rangle
+\langle\widehat{\mathbf{T}}c_{\mathbf{q-p,\downarrow}}(\tau
)c_{\mathbf{p,\uparrow}}(\tau)c_{\mathbf{q-k},\downarrow}^{\dagger}%
(\tau)c_{\mathbf{k},\uparrow}^{\dagger}(\tau^{\prime})\rangle\} \tag{4}%
\end{align}
and
\begin{align}
&  \quad\frac{d}{d\tau}\langle\widehat{\mathbf{T}}[b_{\mathbf{q}}^{\dagger
}(\tau)-\langle b_{0}^{\dagger}\rangle]c_{\mathbf{k+q,\downarrow}}%
(\tau)c_{\mathbf{k},\downarrow}^{\dagger}(\tau^{\prime})\rangle\nonumber\\
&  =[(\omega_{\mathbf{q}}+\nu^{\prime})-\xi_{\mathbf{k+q}}]\langle
\widehat{\mathbf{T}}[b_{\mathbf{q}}^{\dagger}(\tau)-\langle b_{0}^{\dagger
}\rangle]c_{\mathbf{k+q,\downarrow}}(\tau)c_{\mathbf{k},\downarrow}^{\dagger
}(\tau^{\prime})\rangle+(\omega_{\mathbf{q}}+\nu^{\prime})\langle
b_{0}^{\dagger}\rangle\langle\widehat{\mathbf{T}}c_{\mathbf{k+q,\downarrow}%
}(\tau)c_{\mathbf{k},\downarrow}^{\dagger}(\tau^{\prime})\rangle\nonumber\\
&  +U\sum_{\mathbf{p,q}^{\prime}}\langle\widehat{\mathbf{T}}[b_{\mathbf{q}%
}^{\dagger}(\tau)-\langle b_{0}^{\dagger}\rangle]c_{\mathbf{p+q}^{\prime
},\uparrow}^{\dagger}(\tau)c_{\mathbf{k+q}^{\prime}+\mathbf{q,}\downarrow
}(\tau)c_{\mathbf{p,\uparrow}}(\tau)c_{\mathbf{k},\downarrow}^{\dagger}%
(\tau^{\prime})\rangle\nonumber\\
&  +g\sum_{\mathbf{p}}\{\langle\widehat{\mathbf{T}}[b_{\mathbf{q}}^{\dagger
}(\tau)-\langle b_{0}^{\dagger}\rangle]b_{\mathbf{p}}(\tau)c_{\mathbf{p}%
-\mathbf{k-q},\uparrow}^{\dagger}(\tau)c_{\mathbf{k},\downarrow}^{\dagger
}(\tau^{\prime})\rangle-\langle\widehat{\mathbf{T}}c_{\mathbf{q-p,\downarrow}%
}(\tau)c_{\mathbf{p,\uparrow}}(\tau)c_{\mathbf{q-k},\uparrow}^{\dagger}%
(\tau)c_{\mathbf{k},\downarrow}^{\dagger}(\tau^{\prime})\rangle\}. \tag{5}%
\end{align}
At this stage we applied the decoupling between the fermion operators and
boson operators. As before, we assumed the contribution of the interaction
terms of eqs. (4) and (5) (the third and fourth lines) can be incorporated
into the one-particle energies $\xi_{\mathbf{k}}$ and $\nu$. The resulting
forms will be comprehensible and easy to analyze. Thus, eqs. (4) and (5)
became
\begin{equation}
\lbrack\frac{d}{d\tau}+(\omega_{\mathbf{q}}+\nu^{\prime})-\xi_{\mathbf{k-q}%
}]\langle\mathbf{T}[b_{\mathbf{q}}(\tau)-\langle b_{0}\rangle
]c_{-\mathbf{k+q,\downarrow}}^{\dagger}(\tau)c_{\mathbf{k},\uparrow}^{\dagger
}(\tau^{\prime})\rangle\simeq-(\omega_{\mathbf{q}}+\nu^{\prime})\langle
b_{0}\rangle\langle\widehat{\mathbf{T}}c_{-\mathbf{k+q,\downarrow}}^{\dagger
}(\tau)c_{\mathbf{k},\uparrow}^{\dagger}(\tau^{\prime})\rangle\tag{6}%
\end{equation}
and
\begin{equation}
\lbrack\frac{d}{d\tau}-(\omega_{\mathbf{q}}+\nu^{\prime})+\xi_{\mathbf{k+q}%
}]\langle\mathbf{T}[b_{\mathbf{q}}^{\dagger}(\tau)-\langle b_{0}^{\dagger
}\rangle]c_{-\mathbf{k+q,\downarrow}}(\tau)c_{\mathbf{k},\downarrow}^{\dagger
}(\tau^{\prime})\rangle\simeq(\omega_{\mathbf{q}}+\nu^{\prime})\langle
b_{0}\rangle\langle\widehat{\mathbf{T}}c_{-\mathbf{k+q,\downarrow}}%
(\tau)c_{\mathbf{k},\downarrow}^{\dagger}(\tau^{\prime})\rangle. \tag{7}%
\end{equation}
Taking Fourier transform of eqs. (2), (3), (6) and (7) and making
substitutions, we obtained
\begin{equation}
G_{\mathbf{k,\uparrow}}(i\omega_{n})=\frac{i\omega_{n}+\xi_{\mathbf{k}}}%
{\psi(i\omega_{n},\mathbf{k})}, \tag{8}%
\end{equation}
and\
\begin{equation}
F_{\mathbf{k,}\uparrow\downarrow}^{\ast}(i\omega_{n})=\frac{-UF_{\uparrow
\downarrow}^{\ast}+g\phi(i\omega_{n}-\xi_{\mathbf{k}})/(i\omega_{n}%
+\nu^{\prime}-\xi_{\mathbf{k}})}{\psi(i\omega_{n},\mathbf{k})} \tag{9}%
\end{equation}
where $\phi=\langle b_{0}\rangle$ and
\begin{align}
\psi(i\omega_{n},\mathbf{k})  &  =(i\omega_{n}+\xi_{\mathbf{k}})(i\omega
_{n}-\xi_{\mathbf{k}})-[UF_{\uparrow\downarrow}^{\ast}-\frac{g\phi(i\omega
_{n}-\xi_{\mathbf{k}})}{i\omega_{n}+\nu^{\prime}-\xi_{\mathbf{k}}%
}][UF_{\downarrow\uparrow}-\frac{g\phi(i\omega_{n}+\xi_{\mathbf{k}})}%
{i\omega_{n}-\nu^{\prime}+\xi_{\mathbf{k}}}]\nonumber\\
&  =\frac{[(i\omega_{n})^{2}-E_{+,\mathbf{k}}^{2}][(i\omega_{n})^{2}%
-E_{-,\mathbf{k}}^{2}]}{(i\omega_{n})^{2}-(\xi_{\mathbf{k}}-\nu^{\prime})^{2}}
\tag{10}%
\end{align}
with
\begin{equation}
E_{\pm,\mathbf{k}}^{2}=B\pm\{B^{2}-\xi_{\mathbf{k}}^{2}(\xi_{\mathbf{k}}%
-\nu^{\prime})^{2}-U^{2}F_{\uparrow\downarrow}^{\ast2}(\xi_{\mathbf{k}}%
-\nu^{\prime})^{2}-g^{2}\phi^{2}\xi_{\mathbf{k}}{}^{2}+2UF_{\uparrow
\downarrow}^{\ast}g\phi\xi_{\mathbf{k}}(\xi_{\mathbf{k}}-\nu^{\prime})\}^{1/2}
\tag{11}%
\end{equation}
where $B=[\xi_{\mathbf{k}}^{2}+(\xi_{\mathbf{k}}-\nu^{\prime})^{2}%
+(UF_{\uparrow\downarrow}^{\ast}-g\phi)^{2}]/2$. Clearly, the Green's
functions still possess the conventional form of BCS theory. However, we
accounted for the most important ingredient: the coupling between
$F_{\alpha\beta}^{\ast}$ and $\phi$, to insure that we got a valid physical
picture. In fact, as shown below, our results are qualitatively similar to
those of ref. 12 who had considered many Feynman diagrams.

In order to study the resulting physical properties, we followed
Eliashberg[16] to write down the form of the thermodynamic potential
\begin{equation}
\Omega=\sum_{\mathbf{k}}\xi_{\mathbf{k}}+(UF_{\downarrow\uparrow}%
-g\phi)F_{\uparrow\downarrow}^{\ast}-k_{B}T\sum_{n,\mathbf{k}}\ln\psi
(i\omega_{n},\mathbf{k})+(\nu-2\mu)\phi^{2}+k_{B}T\sum_{\mathbf{q}%
}\{1-e^{-[\beta(\omega_{\mathbf{q}}+\nu^{\prime})]}\}. \tag{12}%
\end{equation}
Here, we have kept only the zeroth and first order terms. The relation
$F_{\uparrow\downarrow}^{\ast}=F_{\downarrow\uparrow}$ is also helpful. Now we
take the variational approach.
\begin{align}
N  &  =-\frac{\partial\Omega}{\partial\mu}=\sum_{\mathbf{k}}[1+\frac{\partial
E_{\mathbf{+,k}}}{\partial\mu}\tanh\frac{\beta E_{+,\mathbf{k}}}{2}%
+\frac{\partial E_{-\mathbf{,k}}}{\partial\mu}\tanh\frac{\beta E_{-,\mathbf{k}%
}}{2}-\tanh\frac{\beta(\xi_{\mathbf{k}}-\nu^{\prime})}{2}]\nonumber\\
&  +2\phi^{2}+2\sum_{\mathbf{q}}n_{B}(\omega_{q}), \tag{13}%
\end{align}
and
\begin{equation}
\frac{\partial\Omega}{\partial\phi}=2\phi\nu^{\prime}-gF_{\uparrow\downarrow
}^{\ast}-\sum_{\mathbf{k}}[\frac{\partial E_{\mathbf{+,k}}}{\partial\phi}%
\tanh\frac{\beta E_{+,\mathbf{k}}}{2}+\frac{\partial E_{-\mathbf{,k}}%
}{\partial\phi}\tanh\frac{\beta E_{-,\mathbf{k}}}{2}]=0 \tag{14}%
\end{equation}
where $n_{B}(\omega_{q})=1/[e^{(\beta\omega_{\mathbf{q}}+\nu^{\prime})}-1]$.
Additionally, we have the gap equation
\begin{align}
F_{\uparrow\downarrow}^{\ast}  &  =\sum_{n,\mathbf{k}}F_{\uparrow\downarrow
}^{\ast}(i\omega_{n},\mathbf{k})=\sum_{\mathbf{k}}\frac{UF_{\uparrow
\downarrow}^{\ast}[E_{+,\mathbf{k}}^{2}-(\xi_{\mathbf{k}}-\nu^{\prime}%
)^{2}]-g\phi\lbrack E_{+,\mathbf{k}}^{2}-\xi_{\mathbf{k}}(\xi_{\mathbf{k}}%
-\nu^{\prime})]}{2E_{+,\mathbf{k}}(E_{+,\mathbf{k}}^{2}-E_{-,\mathbf{k}}^{2}%
)}\nonumber\\
&  \qquad\qquad\qquad\quad-\sum_{\mathbf{k}}\frac{UF_{\uparrow\downarrow
}^{\ast}[E_{-,\mathbf{k}}^{2}-(\xi_{\mathbf{k}}-\nu^{\prime})^{2}%
]-g\phi\lbrack E_{-,\mathbf{k}}^{2}-\xi_{\mathbf{k}}(\xi_{\mathbf{k}}%
-\nu^{\prime})]}{2E_{-,\mathbf{k}}(E_{+,\mathbf{k}}^{2}-E_{-,\mathbf{k}}^{2}%
)}. \tag{15}%
\end{align}
With eqs. (13-15) we can solve for three unknowns: $\mu$,\ $\phi$\ and
$F_{\alpha\beta}^{\ast}$.

We showed the results in Figs. 1 by plotting $\phi$ and $F_{\uparrow
\downarrow}^{\ast}$\ versus temperature at detuned energies $\nu=\pm0.5E_{F}%
$\ in the solid and dashed lines respectively. The parameters are the same as
those in ref. 12, i. e., $g=-0.6E_{F}$, $U=0.3E_{F}$ and a cutoff factor
$\exp[-(\varepsilon_{\mathbf{k}}/2E_{F})^{2}]$.\ In Fig. 2 the amplitudes of
the states of phase coherence $F_{\uparrow\downarrow}^{\ast}$\ and $\phi$\ at
\ $k_{B}T=E_{F}/3$ (in the dashed lines) and $k_{B}T=E_{F}/6$ (in the solid
lines) are plotted against the detuned energy. It is remarkable that in Figs.
1 and 2 $F_{\uparrow\downarrow}^{\ast}$\ and $\phi$ coexist in the whole range
and vanish together. There is no crossover. In fact, similar results had been
obtained by previous calculations[11,12] with the static approximation
$\langle b(\tau)\rangle\simeq\phi$ and the resultant relation $\phi
=-gF_{\uparrow\downarrow}^{\ast}/(\nu-2\mu)$ and energy gap $\Delta
=[U+g^{2}/(\nu-2\mu)]F_{\uparrow\downarrow}^{\ast}$. Note that one of the
consequences of the approximation is that $\phi$\ and $F_{\uparrow\downarrow
}^{\ast}$\ are proportional. However, our approach did not have the
proportional relation \textit{a priori}. Therefore, we are able to conclude
that the original concept of BEC-BCS crossover cannot be applied here.

In Fig. 3 we presented $-\operatorname{Im}F_{\uparrow\downarrow}^{\ast}(E)$
versus $E/E_{F}$. On the BCS side $(\nu>0$), the feature of Cooper pair
density of states is clearly shown. It gradually became molecule-like when
$\nu$ became negative.\ The reason, as pointed out by Leggett[2], is that when
the chemical potential becomes negative, the divergence in the spectrum
disappear. Note that in the whole range, the amplitude of BEC is greater than
that of BCS due to the parameters we chose. Yet the spectrum underwent drastic
change. Clearly, the shape of spectrum is not an indication of whether it is
BEC or BCS state.

It should not be surprising that BEC persists deeply into the BCS side of
Feshbach resonance. Although the detuned energy is positive, forming molecules
can still be energetically favorable. It is because most atoms (fermions) have
finite amount of kinetic energy. On the other hand, on the BEC side, it seems
that molecule formation is advantageous. However, BCS state is helped by
having the energy gap. The larger the gap, the lower the energy. For negative
detuned energy, the chemical energy is also negative. As pointed out by
Leggett[2], the gap under this situation is at least $\sqrt{\mu^{2}+\Delta
^{2}}$. The energy gain by forming Cooper pairs grows with the magnitude of
the chemical potential and the detuned energy and thus, the coexistence.
Furthermore, if we simplified our model by considering only the condensed
bosons ($b_{\mathbf{q=0}}$), then the Hamiltonian in eq. (1) would remind one
of the Anderson model[17] or Fano resonance[18]. The eigenstate should be a
mixture of localized state (condensed bosons) and continuum (fermions). This
analogy, though not completely compatible, indicates that the states of phase
coherence should have two components. In conclusion, we pointed out that
either $\langle b(\tau)\rangle$\ or $F_{\alpha\beta}^{\ast}(\tau,\tau)$ should
be time-dependent and our calculation show that BCS state and BEC always coexist.

This work was support in part by NSC of Taiwan, R. O. C., under the contract
NSC 93-2112-M-002-005. The author benefited from the discussions with Lu
Hsin-I and the members of the focused groups of spin-related physics and SCES
of NCTS, Taiwan ROC.

\noindent{\LARGE References}

1. D .M. Eagles, Phys. Rev. \textbf{186}, 456 (1969).

2. A. J. Leggett, ''Modern Trends in the theory of Condensed Matter'', edited
by A. Pekalski and Przystawa, Springer-Verlag, Berlin (1980).

3. P. Nozi\'{e}res and S. Schmitt-Rink, J. Low Temp. Phys. \textbf{59}, 195 (1985).

4. S. Jochim, M. Bartenstein, A. Altmeyer, G. Hendl, S. Riedl, C. Chin, J.
Hecker Denschlag, R. Grimm, Science \textbf{302}, 2101 (2003).

5. Markus Greiner, Cindy A. Regal, and Deborah S. Jin, Nature \textbf{426},
537 (2003).

6. C. A. Regal, M. Greiner and D. S. Jin, Phys. Rev. Lett. \textbf{92}, 040403 (2004).

7. C. Chin, M. Bartenstein, A. Altmeyer, S. Riedl, S. Jochim, J. Hecker
Denschlag, and R. Grimm, Science \textbf{305}, 1128 (2004).

8. M. Greiner, C. A. Regal, and D. S. Jin, Phys. Rev. Lett. \textbf{94},
070403 (2005).

9. H. Feshbach, Ann. Phys. (NY) \textbf{5}, 357 (1958).

10. J. Kinnunen, M. Rodriguez, P. Torma, Science \textbf{305}: 1131 (2004).

11. S. J. J. M. F. Kokkelmans. J. N. Milstein, M. L. Chiofalo, R. Walser and
M. J. Holland, Phys. Rev. \textbf{A65}, 053617 (2002).

12. Y. Ohashi and A. Griffin, Phys. Rev. \textbf{A67}, 063612 (2003).

13. Jelena Stajic, J. N. Milstein, Qijin Chen, M. L. Chiofalo, M. J. Holland
and K. Levin, Phys. Rev. \textbf{A69}, 063610 (2004).

14. M. J. Holland, S. J. J. M. F. Kokkelmans, M. L. Chiofalo and R. Walser,
Phys. Rev. Lett. \textbf{87}, 120406 (2001).

15. A. A. Abrikosov, L. P. Gorkov and I. E. Dzyaloshinski, ''Method of Quantum
Field Theory in Statistical Physics'', translated and edited by Richard A.
Silverman, Dover, NewYork (1977).

16. G. M. Eliashberg, JETP \textbf{16}, 780 (1963).

17. P. W. Anderson, Phys. Rev. \textbf{124}, 41 (1961).

18. U. Fano, Phys. Rev. \textbf{124}, 1866 (1961).\pagebreak 

Figure captions

Fig. 1 $F_{\uparrow\downarrow}^{\ast}$ and $\phi$ versus temperature at
$\nu=0.5E_{F}$ (solid lines) and $\nu=-0.5E_{F}$\ (dashed lines).

Fig. 2 $F_{\uparrow\downarrow}^{\ast}$ and $\phi$ versus the detuned energy
$\nu$ at temperature at $E_{F}/6\ $(solid lines) and $E_{F}/3$ (dashed lines).

Fig. 3 $-\operatorname{Im}F_{\uparrow\downarrow}^{\ast}(E)$, the spectrum
funtion of Cooper pairs versus $E/E_{F}$ at different detuned
energies.\pagebreak
%TCIMACRO{\FRAME{ftbpF}{4.9684in}{7.0257in}{0pt}{}{}{colddpm05.eps}%
%{\special{ language "Scientific Word";  type "GRAPHIC";
%maintain-aspect-ratio TRUE;  display "USEDEF";  valid_file "F";
%width 4.9684in;  height 7.0257in;  depth 0pt;  original-width 8.2434in;
%original-height 11.6836in;  cropleft "0";  croptop "1";  cropright "1";
%cropbottom "0";  filename 'colddpm05.EPS';file-properties "XNPEU";}}}%
%BeginExpansion
\begin{figure}
[ptb]
\begin{center}
\includegraphics[
natheight=11.683600in,
natwidth=8.243400in,
height=7.0257in,
width=4.9684in
]%
{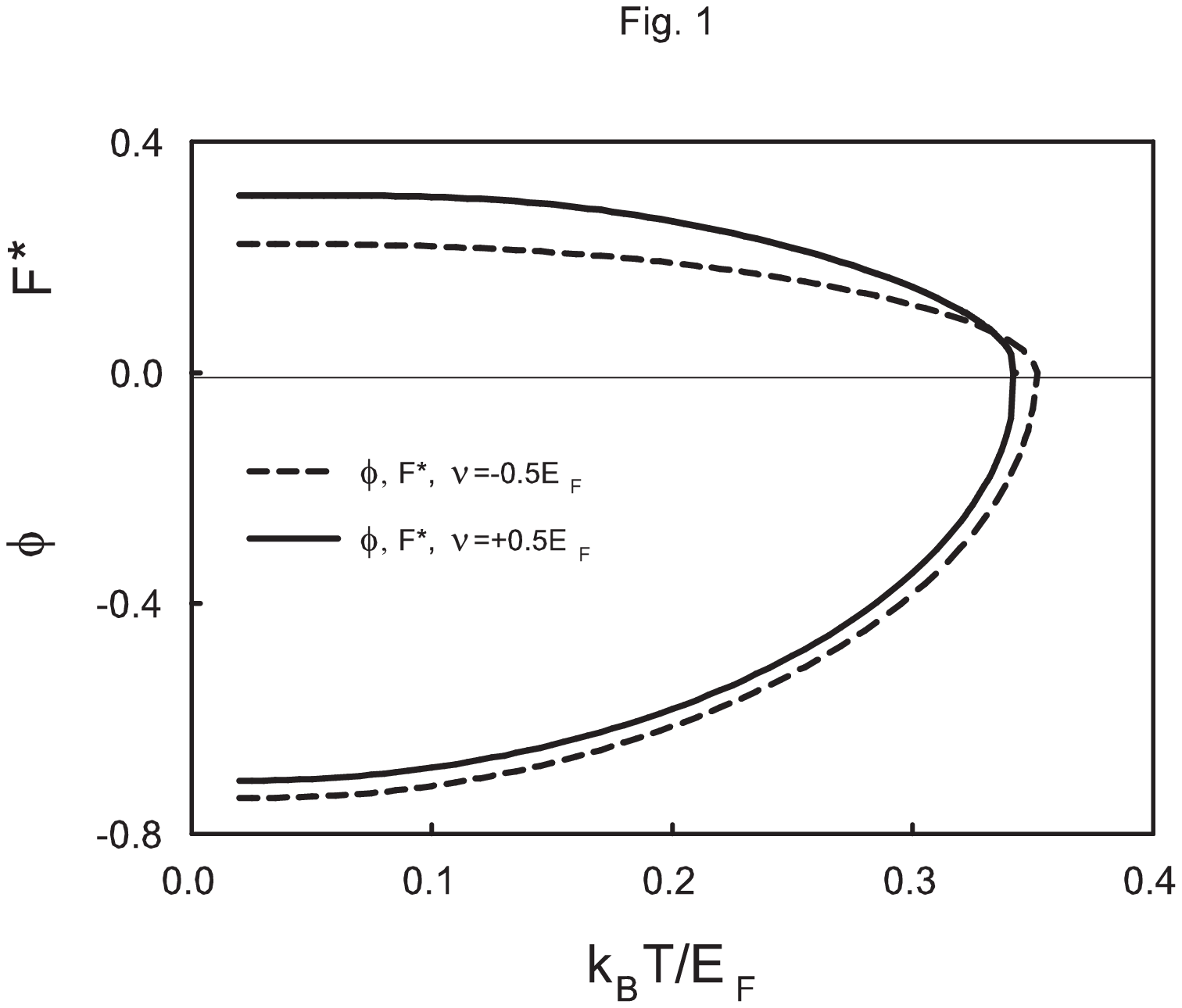}%
\end{center}
\end{figure}
%EndExpansion
\pagebreak
%TCIMACRO{\FRAME{ftbpF}{5.0211in}{7.0984in}{0pt}{}{}{coldk01633.eps}%
%{\special{ language "Scientific Word";  type "GRAPHIC";
%maintain-aspect-ratio TRUE;  display "USEDEF";  valid_file "F";
%width 5.0211in;  height 7.0984in;  depth 0pt;  original-width 8.2434in;
%original-height 11.6836in;  cropleft "0";  croptop "1";  cropright "1";
%cropbottom "0";  filename 'coldk01633.EPS';file-properties "XNPEU";}}}%
%BeginExpansion
\begin{figure}
[ptbptb]
\begin{center}
\includegraphics[
natheight=11.683600in,
natwidth=8.243400in,
height=7.0984in,
width=5.0211in
]%
{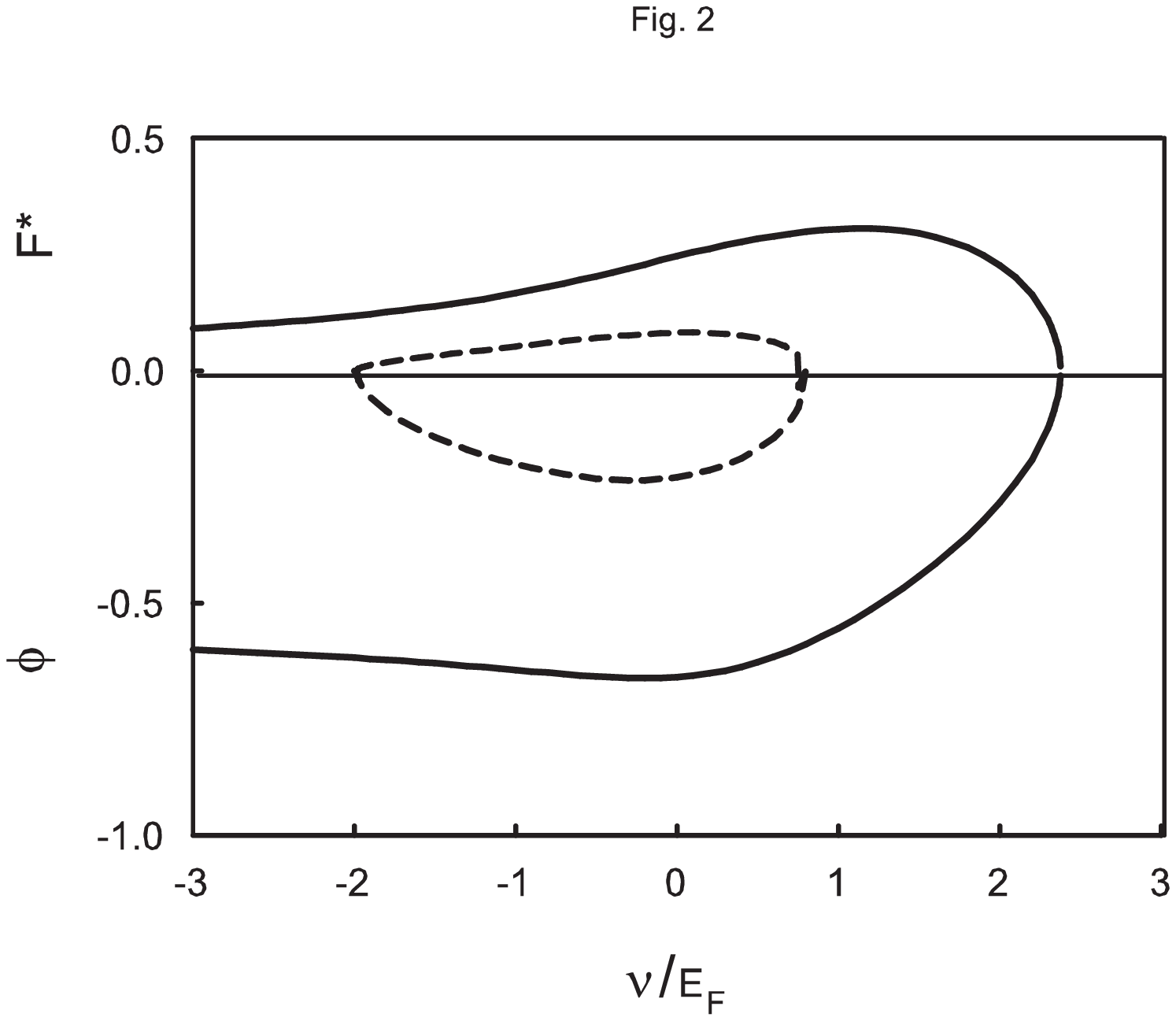}%
\end{center}
\end{figure}
%EndExpansion
\pagebreak
%TCIMACRO{\FRAME{ftbpF}{5.0073in}{7.0802in}{0pt}{}{}{coldfs.eps}%
%{\special{ language "Scientific Word";  type "GRAPHIC";
%maintain-aspect-ratio TRUE;  display "USEDEF";  valid_file "F";
%width 5.0073in;  height 7.0802in;  depth 0pt;  original-width 8.2434in;
%original-height 11.6836in;  cropleft "0";  croptop "1";  cropright "1";
%cropbottom "0";  filename 'coldfs.EPS';file-properties "XNPEU";}}}%
%BeginExpansion
\begin{figure}
[ptbptbptb]
\begin{center}
\includegraphics[
natheight=11.683600in,
natwidth=8.243400in,
height=7.0802in,
width=5.0073in
]%
{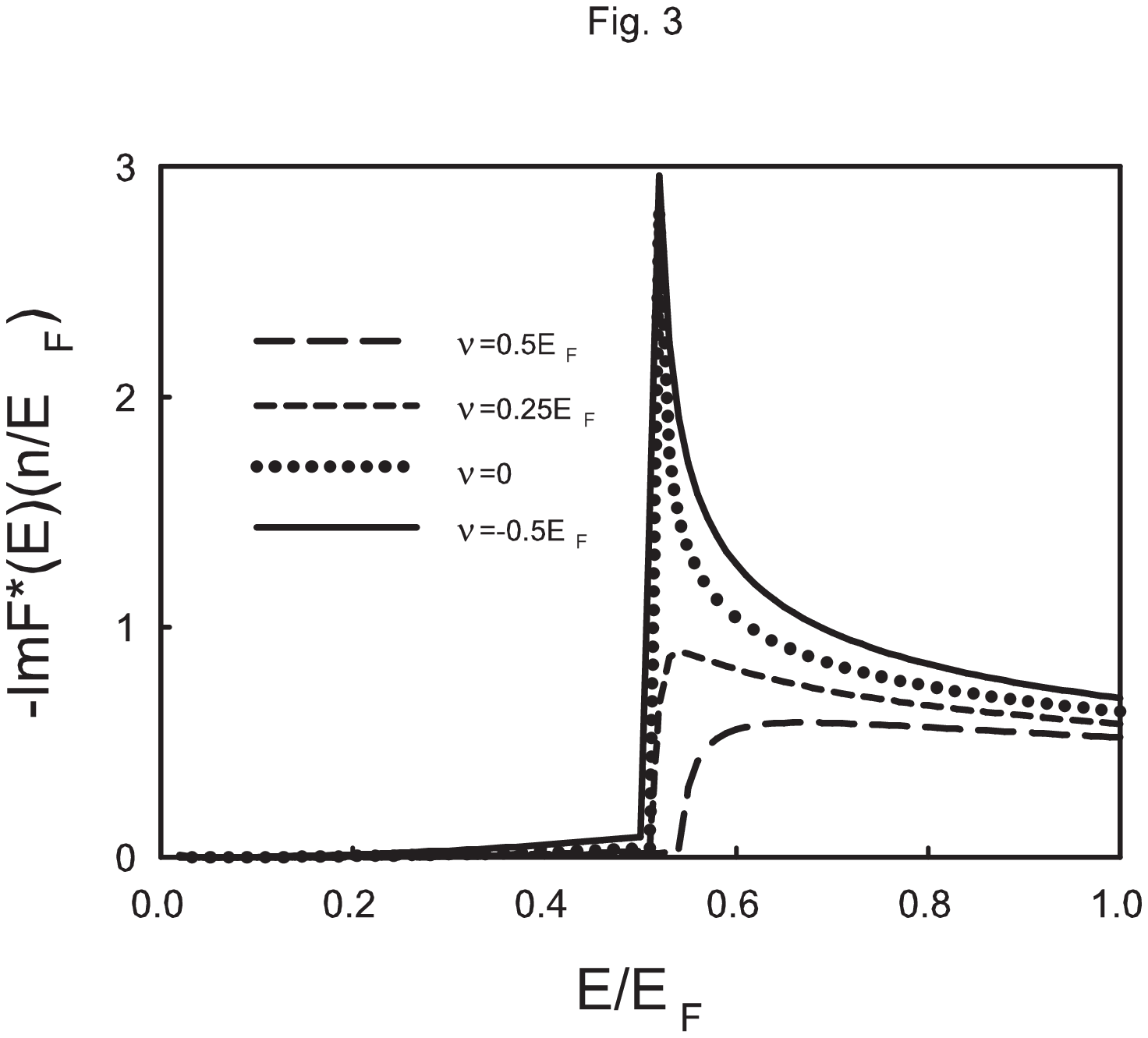}%
\end{center}
\end{figure}
%EndExpansion
\end{document}